\documentclass[prd,preprintnumbers,amsmath,amssymb, nofootinbib]{revtex4}
\usepackage{amsmath,amssymb,graphicx}
\usepackage{epsf}
\usepackage{epstopdf}
\usepackage {amssymb}
\newcommand{\nc}{\newcommand}
\nc{\ba}{\begin{eqnarray}}
\nc{\ea}{\end{eqnarray}}
\newcommand\be{\begin{equation}}
\newcommand\ee{\end{equation}}

\def\bal#1\eal{\begin{align}#1\end{align}}

\nc{\bk}{{\mathbf{k}}}
\nc{\bq}{{\mathbf{q}}}
\nc{\bp}{{\mathbf{p}}}
\nc{\bn}{{\mathbf{n}}}
\nc{\calR}{{\cal R}}
\nc{\calP}{{\cal P}}

\input{epsf.sty}

\begin{document}


\title{Anisotropic Inflation with the non-Vacuum Initial State  }
\date{\today}

\author{Razieh Emami$^{1}$}
\email{emami-AT-ipm.ir}
\author{Hassan Firouzjahi$^{2}$}
\email{firouz-AT-mail.ipm.ir}
\author{Moslem Zarei $^{3, 1} $}
\email{m.zarei-AT-cc.iut.ac.ir}

\affiliation{$^1$School of Physics, Institute for Research in
Fundamental Sciences (IPM), P.~O.~Box 19395-5531,
Tehran, Iran}

\affiliation{$^2$School of Astronomy, Institute for Research in
Fundamental Sciences (IPM),
P.~O.~Box 19395-5531,
Tehran, Iran}

\affiliation{ $ ^3$ Department of Physics, Isfahan University of Technology, Isfahan 84156-83111, Iran }

\begin{abstract}
\vspace{0.3cm}

In this work we study models of anisotropic inflation with the generalized non-vacuum initial states for the inflaton field and the gauge field. The effects of non Bunch-Davies initial condition on the anisotropic power spectrum and bispectrum are calculated. We show that the non Bunch-Davies initial state can help to reduce the fine-tuning on the anisotropic power spectrum while reducing the level of anisotropic bispectrum. 

\vspace{0.3cm}

\end{abstract}

\date\today

\maketitle


\section{Introduction}

Statistical isotropy of the cosmological background is a fundamental assumption in standard cosmology which is well-supported by different cosmological observations at different scales and red-shifts. The principle of statistical isotropy on large scale is also motivated from the Copernicus point of view in which there is no preferred direction or reference point in the universe. However, there are indications for the violation of statistical isotropy in recent cosmological observations such as  WMAP  \cite{Bennett:2012zja,Hinshaw:2012fq}
and  PLANCK \cite{Ade:2013lta}, for a detailed review see \cite{Ade:2013nlj}. 

Anisotropic inflation is an interesting mechanism for generating statistical anisotropies in CMB map. In order to break the conformal invariance a $U(1)$ gauge field is non-minimally coupled to inflaton field such that the gauge field survives the exponential expansion.  A simple and theoretically well-motivated model of anisotropic inflation is based on the theory with the Lagrangian  $f(\phi)^2 F_{\mu}F^{\mu}/4$ in which $\phi$ is the inflaton field, 
$F_{\mu \nu}$ is the gauge field strength and $f(\phi)$ is the  inverse of the gauge kinetic coupling \cite{Turner:1987bw, Ratra:1991bn}, for a review see \cite{Soda:2012zm, Maleknejad:2012fw} and the references therein.  An interesting realization of anisotropic inflation was put forward in  \cite{Watanabe:2009ct}, see also 
\cite{Ohashi:2013pca, Ohashi:2013mka, Emami:2010rm, Thorsrud:2012mu}, 
in which  with an appropriate choice of the coupling $f(\phi)$ one can break the conformal invariance such that an attractor mechanism is generated for the gauge field dynamics. During the attractor phase  the fraction of the 
gauge field energy density to the total energy density is nearly constant,  at the order of slow-roll parameters \cite{Watanabe:2009ct}.  As a result, statistical anisotropies can be generated which are small but can be observable.  

Quadrupolar  asymmetry is a generic predictions of models of anisotropic inflation. The anisotropies induced on primordial curvature perturbation power spectrum $\calP_{\calR }$
has the form 
\ba
\label{g-def}
\calP_\calR = \calP_{\calR }^{(0)} \left( 1+ g_* (\hat {\bf n}. \hat{\bf k})^2 \right)
\ea 
in which $\calP_{\calR }^{(0)}$ is the isotropic power spectrum,  
$\hat{\bf k}$ represents the momentum direction in Fourier space while $\hat {\bf n}$
is the preferred direction in the sky. In this view, $g_{*}$ measures the amplitude of the quadrupole asymmetry. There are strong observational constraints on  value of $g_*$. 
At 95 \% CL the  PLANCK data implies $ -0.05< g_* < 0.05$ and
$-0.36 < g_* < 0.36$ from $L=0, L=2$ modes respectively. On the other hand, 
the constrains from PLANCK data were further investigated in  \cite{Kim:2013gka}, yielding $g_* =0.002 \pm 0.016$ at 68\% CL.  Despite the strong observational constrains on  quadrupolar asymmetry the the possibility of having a  quadrupolar asymmetry  in primordial power spectrum is interesting  theoretically.

One peculiar feature of models of anisotropic inflation is that $g_*$ quadratically scales with 
$N$, the total number of e-foldings \cite{Dulaney:2010sq, Gumrukcuoglu:2010yc, Watanabe:2010fh, Bartolo:2012sd, 
Funakoshi:2012ym, Yamamoto:2012sq, Emami:2013bk, 
Shiraishi:2013vja, Abolhasani:2013zya, 
Abolhasani:2013bpa}.
This indicates that if the duration of anisotropic inflation in the attractor regimes last long enough, much longer than the minimum 60  or so e-foldings to solve the flatness the horizon problem, then too much anisotropies are generated. Physically, this originates from the fact that once the gauge field excitations leave the horizon they become super-horizon and create an effective anisotropic background \cite{Bartolo:2012sd}. The induced infra-red (IR) anisotropies generated from the gauge field fluctuations can 
accumulate to destroy the background slow-roll inflation. This is in contrast to  scalar field fluctuations in which the super-horizon scalar field perturbations carry no preferred directions and no IR anisotropies are generated.  

In order not to produce too much IR anisotropies in models of anisotropic inflation, one has to chose the minimum value of $N$ required to solve the flatness and the horizon problem. In this situation one has to worry about the initial quantum state of the the universe. In conventional models of inflation one usually assumes that inflation lasts very long in the past so for the cosmological scales, i.e. CMB scales perturbations, one can safely assume the vacuum or the Bunch-Davies (BD) initial state. This is motivated from the fact that the BD vacuum has the minimum energy so if inflation continues long enough in the past then the universe eventually ends up in a BD vacuum state.  Now in models of anisotropic inflation with a finite value of $N$ one has to consider a generalized initial state which may not be the vacuum state. In other words, the inflationary universe did not have much time to settles down to a BD vacuum.  With this motivation in mind,  in this paper we would like to investigate the effects of non-BD initial conditions for the anisotropic power spectrum and bispectrum in models of anisotropic inflation. For earlier works on non-BD effects on anisotropies see also \cite{Chen:2013tna, Chen:2013eaa}. 
Indeed, the issue of non-BD initial state have been studied extensively in recent literature \cite{Agullo:2010ws, Ganc:2011dy, Chialva:2011hc, Creminelli:2011rh,
Ganc:2012ae, Agullo:2012cs, Agarwal:2012mq, Ashoorioon:2010xg, Chen:2006nt,Holman:2007na, Meerburg:2009ys, Ashoorioon:2013eia, Kundu:2013gha, Brahma:2013rua, 
Bahrami:2013isa, Gong:2013yvl}.  In particular, a non-BD initial condition yields 
large local-type non-Gaussianity in single field models of inflation. This is one of the known mechanism to violate the celebrated Maldacena's consistency condition \cite{Maldacena:2002vr} for the single field inflation models\footnote{ For another known method of violating the Maldacena's consistency condition see  \cite{Namjoo:2012aa, Chen:2013aj, Huang:2013lda, Chen:2013kta} in which there are some non-attractor phase of inflation at the early stage of inflation. } which can have important observational implications.

\section{Anisotropic Inflation}

In this section we review anisotropic inflation briefly following the analysis of   \cite{Watanabe:2010fh} and \cite{Emami:2013bk}. First we present the background 
while the perturbations are presented in next sub-sections.

\subsection{Background}
\label{Background }

The model consist of an inflaton field in the presence of a $U(1)$ gauge field. 
The action is given by
\ba
\label{action}
S= \int d^4 x \sqrt{-g} \left [ \frac{M_P^2}{2}
R -  \frac{1}{2} \partial_\mu \phi\partial^\mu \phi - \frac{f^2(\phi)}{4}F_{\mu \nu} F^{\mu \nu}  - V(\phi) \right]
\ea
in which $\phi$ is the inflaton field and as usual $F_{\mu\nu} = \partial_\mu A_\nu - \partial_\nu A_\mu $ represents  the field strength obtained from the $U(1)$ gauge field $A_\mu$.
It is assumed  that the gauge field has a non-zero background value along the x-direction so $A_\mu =(0,A_{x}(t),0,0)$. This breaks the isotropy of the space-time and the background  has the form of  Bianchi I Universe with the metric 
\ba
ds^2 &=& - dt^2 + e^{2\alpha(t)}\left( e^{-4\sigma(t)}d x^2 +e^{2\sigma(t)}(d y^2 +d z^2) \right)  \nonumber\\
&=&   - dt^2 + a(t)^2 dx^2 + b(t)^2 (dy^2 + dz^2) \, .
\ea
With  this space-time metric $H \equiv \dot \alpha$ represents the average Hubble expansion rate while $H_a \equiv \dot a/a$ and $H_b \equiv \dot b/b$ measure the expansion rates along the spatial directions $x$ and $y$. We also define $\dot \sigma/H \equiv (H_b - H_a)/ H$ as a measure of anisotropic expansion.

The background fields equations are given by
\ba
\label{back-A-eq}
\partial_t{\left(  f^2(\phi) e^{\alpha + 4 \sigma} \dot A_x        \right)}& =& 0 \\
\label{back-rho-eq}
\ddot\phi+3\dot \alpha\dot \phi+ V_\phi
-f(\phi)f_{,\phi}(\phi)\dot A_x^2   e^{-2\alpha+4\sigma}&=&0  \\
\label{Ein1-eq}
\frac{1}{2}\dot
\phi^2+V(\phi)+    \frac{1}{2}f^2(\phi)\dot
A_x^2  e^{-2\alpha+4\sigma}
&=&
3 M_P^2 \left(   \dot \alpha^2-\dot \sigma^2 \right)  \\
\label{Ein2-eq}
V(\phi)+   \frac{1}{6}f^2(\phi)\dot
A_x^2  e^{-2\alpha+4\sigma}
&=& M_P^2 \left( \ddot \alpha    + 3 \dot \alpha^2 \right)  \\
\label{anisotropy-eq}
\frac{1}{3}f^2(\phi)\dot A_x^2   e^{-2\alpha+4\sigma}
&=& M_P^2\left( 3\dot \alpha \dot \sigma+ \ddot \sigma      \right)\, ,
\ea
where a dot represents the derivative with respect to cosmic time $t$.

Fortunately the Maxwell equation  for $A_{x}$, Eq. (\ref{back-A-eq}), is easily solved to yield
\ba
\label{gaugefield}
\dot{A_{x}}= f(\phi)^{-2}e^{-\alpha(t)-4\sigma(t)}p_{A} \, ,
\ea
in which  $p_{A}$ is a constant of integration.

In general the above system of equations is too complicated to be solved analytically. On the physical ground, we are interested in the small anisotropy limit, $|\dot \sigma/H | \ll 1 $. As a result one expects that  the background expansion is mainly driven by the isotropic potential as in conventional models of inflation.  In order for the anisotropy to be small we demand that
the ratio of the  electric field energy density associated with the gauge field  to the total potential  to be small. Denoting this ratio by $R$ we therefore look for 
 $R \ll 1$ in which
\ba
\label{R-def}
R \equiv \frac{\dot A^2 f(\phi)^2 e^{-2 \alpha}}{2 V} \, .
\ea

During the attractor regime  the anisotropy is small, $R\ll 1$, so the Hubble expansion rate in modified Friedmann equation (\ref{Ein1-eq}) is mainly determined by the potential term. However, the back-reactions of the gauge field on the inflaton field can not be neglected. It 
induces an effective mass for the inflaton field as shown by the last term in Eq. (\ref{back-rho-eq}). As a result, the dynamics of the inflaton field is different than the conventional isotropic models.   Choosing  an appropriate form of the gauge kinetic coupling $f(\phi)$ the system reaches an attractor solution in which $R$ reaches a sub-dominant but nearly constant value \cite{Watanabe:2009ct}.  

In order to obtain a nearly constant value of $R$ one requires $f(\phi) \propto a^{n}$ with $n\simeq -2$. In general, the background expansion is given by
\ba
\label{a-scale}
a \propto \exp \left[ - \int d \phi \frac{V}{ V_\phi} \right] \, .
\ea
Therefore, if we consider 
\ba
\label{f-scale}
f \propto \exp \left[ -n  \int d \phi \frac{V}{V_\phi} \right]
\ea
then $f$  scales as $f \propto a^{n}$. To find the exact form of $f(\phi)$ one has to specify the potential $V(\phi)$.  For the chaotic potential $V= \frac{1}{2} m^2 \phi^2$ one obtains 
\ba
\label{Chaoticpotential}
f(\phi) = \exp {\left( \frac{c\phi^2}{2 M_P^2}  \right)} \, ,
\ea
with $c$ a constant such that $c\ge 1$. In order to obtain small anisotropy we require to tune $c$ very close to unity.   

One can also express $f$ in terms of time or the scale factor $a(t) = e^{\alpha}$ as 
\ba
\label{f-form}
f= \left( \frac{a}{a_f} \right)^{-2 c}  \simeq \left( \frac{\eta}{\eta_e}\right)^{2c} \, ,
\ea
in which $\eta $ is the conformal time related to the cosmic time via $d\eta = dt/a(t)$ and 
$a_e$ and $\eta_e$ represent the values of the scale factor and the conformal time at the end of inflation.

As shown in \cite{Watanabe:2009ct} during the attractor regime  $R$ scales like the slow-roll parameter  given by
\ba
\label{R-app}
R = \frac{c-1}{2c}\epsilon_{H} = \frac{1}{2}I\epsilon \, ,
\ea
in which  the anisotropy parameter $I$  is defined via $I\equiv\frac{c-1}{c}$ and $\epsilon \equiv \dot H/H$ is the slow-roll parameter. In addition,  the anisotropic expansion is given by
\ba
\label{anis-H}
\frac{\dot \sigma}{\dot \alpha} \simeq \frac{I \epsilon}{3}\, .
\ea

\subsection{Perturbations}

Here we review the perturbations in anisotropic inflation backgrounds. We follow the convention of \cite{Emami:2013bk}.

Because we turn on  the background gauge field  along the $x$-direction, the three-dimensional rotation symmetry is broken into a two-dimensional rotation symmetry in $y-z$ plane. Therefore, in order to classify the perturbations in this setup one should  look at the transformation properties of the physical fields under the rotation in $y-z$ plane.
As mentioned in \cite{Watanabe:2010fh, Emami:2013bk} the metric and matter perturbations are classified into the scalar and vector perturbations. In addition,  it is important to note that there are no tensor perturbations in two dimensions.

The general form of the metric and matter perturbations and their transformation properties under a general coordinate transformation have been studied in \cite{Emami:2013bk}, see also \cite{Watanabe:2010fh}. One crucial conclusion from these studies is that the dominant contributions in curvature perturbation anisotropies are generated from the matter sector perturbations and the contribution from the metric sector perturbations are sub-leading. This is specifically demonstrated for the power spectrum anisotropies in \cite{ Emami:2013bk} and implicitly for bispectrum anisotropies in \cite{Bartolo:2012sd} and \cite{Abolhasani:2013zya}.
As a result, in the analysis below we neglect the metric perturbations and work with the background Bianchi I universe.

Our system enjoys a subset of two-dimensional symmetry in $y-z$ plane so we choose the 
coordinate system such that the Fourier wave number has the following form
\ba
\label{ab}
\bk= (k_{x} , k_{y} , 0) ~~~,~~~ k_{x} \equiv k \cos{\theta} ~~~,~~~k_{y} \equiv\frac{b}{a} k \sin{\theta}~~.
\ea
In this coordinate system, the  scalar and the vector perturbations of the matter sector, $\delta A_\mu^{(S)}$  and $\delta A_\mu^{(V)}$, are
\ba
\delta A_\mu^{(S)} = (\delta A_0, \delta A_x, \partial_y M, 0) \quad \quad , \quad \quad
\delta A_\mu^{(V)} =(0, 0, 0, D) \, .
\ea
One can check that the scalar and the vector perturbations do not mix with each other in quadratic action and one can look at their excitations and propagations separately. In this work we are interested in anisotropies generated in curvature perturbation power spectrum and bi-spectrum so we do not consider the vector excitations any further.

\subsection{The Perturbations Actions}

In order to find the normalized wave-function we need the action of the free theory. In addition, to calculate the anisotropies generated in power spectrum and bispectrum we need  the second order exchange vertex interactions.  The full second order action is 
given in Eq. (B1) in \cite{Emami:2013bk}. Neglecting the metric perturbations as mentioned above yields 
\ba
\label{S2-scalar-k}
S_2 = &&\int d \eta d^3 k \left[ \frac{b^2}{2} |\delta \phi'|^2
 - \frac{b^2}{2} k_x^2 | \delta \phi|^2 - \frac{a^2}{2} k_y^2 |\delta \phi|^2
+ \frac{b^2}{2 a^2} f^2  |\delta A_1'|^2 +   \frac{b^2}{2 a^2} f^2  k_x^2 |\delta A_0|^2
+ \frac{f^2}{2} k_y^2| \delta A_0 |^2
\right. \nonumber \\ &&\left.
- i k_x \frac{b^2 f^2}{2 a^2}  (\delta A_1'^* \delta A_0 - \delta A_1' \delta A_0^*)
 + \frac{f^2}{2} k_y^2 |M'|^2  - \frac{f^2}{2} k_y^2 (M'^* \delta A_0 + M' \delta A_0^*) - \frac{f^2}{2} k_y^2 |\delta A_1 |^2
- \frac{f^2}{2} k_x^2 k_y^2 |M |^2 
\right. \nonumber \\ &&\left.
+ i k_x \frac{f^2}{2} k_y^2 (\delta A_1^* M - \delta A_1 M^*)
+ \frac{b^2 f f_{, \phi}}{a^2}  A_x' (\delta A_1'^* \delta \phi + \delta A_1' \delta \phi^*)
 + i k_x \frac{b^2 f f_{, \phi}}{a^2}  A_x' (\delta A_0^* \delta \phi - \delta A_0 \delta \phi^*)
 - \frac{f^2}{2} k_y^2 |\delta A_1 |^2\right. \nonumber \\ &&\left.
+ \frac{b^2  f_{, \phi}^2}{2 a^2} A_x'^2 |\delta \phi |^2 +  \frac{b^2 f f_{, \phi \phi}}{2 a^2} A_x'^2 |\delta \phi |^2 - \frac{a^2 b^2}{2} V_{, \phi \phi} |\delta \phi |^2
\right] \, .
\ea
After integrating out the non-dynamical field $\delta A_0$, one obtains an additional contributions into the action.  Denoting this additional contribution by 
$\Delta L_{\delta A_0}$ we have 
\ba
\Delta L_{\delta A_0}  = -\frac{2 a^2}{ b^2 f^2 k^2} \left| i k_x \frac{b^2 f^2}{2 a^2} \delta A_1' - \frac{f^2 k_y^2}{2} M' + i k_x \frac{b^2 f f'}{a^2} A_x' \delta \phi
\right|^2 \, .
\ea
Adding this into the action, the total action for the dynamical fields $\delta \phi, M$ and $\delta A_1$ becomes
\ba
L_2= L_{\delta \phi \delta \phi}^{(0)} + L_{\delta A_1 \delta A_1}^{(0)} + L_{MM}^{(0)} + L_{\delta \phi \delta A_1}^{(0)}+ L_{\delta \phi M }^{(0)} + L_{M \delta A_1}^{(0)} + \Delta L_{\delta A_0}
\ea
in which $ L_{\delta \phi \delta \phi}^{(0)}, L_{\delta A_1 \delta A_1}^{(0)}$ and so on  represent the quadratic Lagrangians directly coming from the action Eq. (\ref{S2-scalar-k}) without taking into account the additional contributions from $\Delta L_{\delta A_0}$.  

More specifically, 
\ba
L_{\delta \phi \delta \phi}^{(0)}&=&  \frac{b^2}{2} |\delta \phi'|^2
+ \left(  - \frac{b^2}{2} k_x^2 - \frac{a^2}{2} k_y^2   - \frac{a^2 b^2}{2} V_{, \phi \phi}
+  \frac{b^2 f f_{, \phi \phi}}{2 a^2} A_x'^2  + \frac{b^2  f_{, \phi}^2}{2 a^2} A_x'^2
\right)  | \delta \phi|^2
\\
L_{MM}^{(0)} &=&  \frac{f^2}{2} k_y^2 |M'|^2 - \frac{f^2}{2} k_x^2 k_y^2 |M |^2
\\
L_{A_1 A_1}^{(0)} &=& \frac{b^2}{2 a^2} f^2  |\delta A_1'|^2 - \frac{f^2}{2} k_y^2 |\delta A_1 |^2 
\\
L_{\delta \phi \delta A_1}^{(0)} &=&  \frac{b^2 f f_{, \phi}}{a^2}  A_x' (\delta A_1'^* \delta \phi + \delta A_1' \delta \phi^*)
\\
L_{\delta \phi M}^{(0)} &=& 0
\\
L_{M A_1}^{(0)} &=& i k_x \frac{f^2}{2} k_y^2 \, \delta A_1^* M
\ea

Now let us define the transverse mode $D_1$ and the longitudinal mode $D_2$ as follows \cite{Emami:2013bk}
\ba
\label{D1D2}
D_1 &\equiv & \delta A_1 - i k \cos \theta M \\
D_2 & \equiv & \cos \theta \delta A_1 + i k \sin^2 \theta M
\ea
One can easily check that the total action becomes
\ba
\label{L2-final}
L_2 = L_{\phi \phi} + L_{D_1 D_1} + L_{\delta \phi \delta D_1}
\ea
in which
\ba
L_{\delta \phi \delta \phi} &=&  \frac{b^2}{2} |\delta \phi'|^2
+ \left(  - \frac{b^2}{2} k_x^2 - \frac{a^2}{2} k_y^2   - \frac{a^2 b^2}{2} V_{, \phi \phi}
+  \frac{b^2 f f_{, \phi \phi}}{2 a^2} A_x'^2
+ \frac{b^2 f'^2}{a^2}  \sin^2 \theta\right)  | \delta \phi|^2
\\
L_{D_1 D_1} &=& \frac{b^2 f^2}{2 a^2} {\sin^2 \theta} (  | D_1'|^2  - k^2  |D_1|^2 )
\\
\label{L-rho-D1}
L_{\delta \phi D_1} &=& \frac{\sqrt{6 I}}{\eta} \frac{b^2 f}{a} (D_1'^* \delta \phi + D_1' \delta \phi^*) \sin^2 \theta
\ea
The advantage of the decomposition into the transverse and longitudinal modes is that we can directly see that the longitudinal mode $D_2$ is not physical and it is not excited. Also note that we have not imposed any gauge on the gauge field excitations so
our analysis for the gauge field fluctuations are performed gauge invariantly. Alternatively,  one can impose a gauge from the start, say the Coulomb-radiation gauge,  and calculate the action for the remaining degrees of freedom. But our gauge-invariant method has the advantage that the non-physical nature of the longitudinal mode becomes specific.

The canonically normalized fields obtained from the actions $L_{\delta \phi \delta \phi}$
and $L_{D_1 D_1}$ are  $\overline \delta \phi $ and $\overline D_1$ in which 
\ba
\overline {\delta \phi_k}  &\equiv& b \delta \phi_k  \equiv u_k\\
\overline {D_{1 \, k}} & \equiv & \frac{b f}{a} \sin \theta D_{1\, k} \equiv \frac{b f}{a} \sin \theta v_k
\ea

\section{The Effects of non Bunch-Davies Initial Conditions}

In the previous section we have obtained the free theory and the exchange vertex interactions between $\delta \phi$ and $D_1$. Now we study the effects of non-BD initial conditions on anisotropic power spectrum and bispectrum.

The free actions for $\overline {\delta \phi}$ and $\overline {D_1}$ describes massless scale-invariant fields fluctuations in a near dS background. The wavefunction of the canonically normalized fields has the following profile
\ba
\label{nBD-mode}
{\cal M}_k= \frac{\alpha_k^{I}}{\sqrt{2 k}} e^{- i k \eta } ( 1- \frac{i}{k \eta})
+ \frac{\beta_k^{I}}{\sqrt{2 k}} e^{ i k \eta } ( 1+ \frac{i}{k \eta})   
\quad ,  \quad {\cal M}_k = \{  \overline {\delta \phi_k},  \overline {D_{1 \, k}} \} 
\ea
in which 
the index $I$ represents the inflaton field or the gauge field, $I= \{ \phi, A \}$.
Note that in the BD case with the Minkowski vacuum, $\alpha_k^{I}=1$ and $\beta_k^{I}=0$.
The decomposition of modes in Eq. (\ref{nBD-mode}) represents the Bogoliubov transformation of the Minkowski vacuum containing both the positive frequency, 
$e^{- i k \eta }$, and the negative frequency $e^{ i k \eta }$. Note that we have allowed for the possibility that the gauge field and the inflaton field fluctuations can have different Bogoliubov  coefficients, i.e. $\alpha_k^{( \phi)} \neq \alpha_k^{ (A) }$ and
$\beta_k^{( \phi)} \neq \beta_k^{(A)}$. This may come from the fact that the gauge field and the inflaton field may experience  different histories in the past inflationary era. For example, there may be features in the inflation past history which can affect the gauge field and the inflaton field differently. For example, there may exist particle creations or phase transition only  in one sector which can effectively back-react on the corresponding sector, without affecting the other sector.  

The Bogoliubov  coefficients $\alpha_k^{I}$ and $\beta_k^{I}$ are subject to the normalization condition
\ba
\label{morm}
| \alpha_k^I|^2 - | \beta_k^I|^2 =1 \, .
\ea
With this normalization condition, we parameterize the Bogoliubov  coefficients as
\ba
\alpha_k^I = \cosh \chi_k^I \, e^{i \omega_k^I} \quad , \quad
\beta_k^I  = \sinh \chi_k^I   \, e^{i \Omega_k^I} \, .
 \ea
in which $\chi_k^I, \omega_k^I$ and $\Omega_k^I$ are three real variables. The above decomposition will be used in the following anisotropy analysis. For the future reference it is also helpful to define the relative phase $\varphi_k^I$ between $\alpha_k^I$ and $ \beta_k^I $ as 
\ba
\label{varphi}
 \varphi_k^I \equiv \omega_k^I - \Omega_k^I \, .
\ea

The natural question is what the physical constrains on the Bogoliubov  coefficients $\alpha_k^{I}$ and $\beta_k^I$ are. There are few conservative constraints which should be implemented when considering non-BD initial condition \cite{Holman:2007na}. The first condition is that the total energy density associate with the non-BD fluctuations to be finite. Suppose we interpret the non-BD initial state as the state in which there are particle excitations with the number density 
\ba
{\cal N}_k \equiv | \beta_k^I|^2 \, ,
\ea
so the number density of the quanta in the proper unit volume is $ | \beta_k|^2 d^3 k/( 2 \pi a(t))^3 $. Summing the energy associated with these modes, we require their contribution in energy density to remain finite so $\int d^3 k \, k\, {\cal N}_k$ converges. This requires that 
${\cal N}_k = {\cal O}(1/k^{4+\delta)}$ with $\delta >0$ in the UV region. This can be interpreted as the renormalizability condition. The second, stronger constraint is that the back-reaction from the non-BD excited states do not halt inflation. This requires that 
\ba
\int d^3 k k\, {\cal N}_k \lesssim M_P^2 H^2 \, .
\ea
Finally, one should make sure that the non-BD fluctuations do not induce too strong scale-dependence in curvature perturbation power spectrum. The change in the spectral index induced from non-BD fluctuations, $\delta n_s$, is \cite{Ganc:2011dy} \, .
\ba
\delta n_s = \frac{d \ln(1 + 2 {\cal N}_k)}{d \ln k}
\ea
From the PLANCK and WMAP constraints we require $n_s \simeq 0.97$ so the change in 
$\delta n_s$ can be at most at the order of few percent.

With this discussion in mind, one may try to consider a phenomenological model for the non-BD effects. One phenomenological modeling is \cite{Holman:2007na}
\ba
\beta_k^I = {\beta}_{ 0} e^{-k^2/a(\tau_0)^2 M^2 }
\ea
in which $M$ is a UV cut-off of the theory. The condition that the back-reaction from the exited non-BD states do not destroy the slow-roll inflation implies \cite{Holman:2007na}
\ba
\label{beta0}
\beta_0 \leq \sqrt{\epsilon} \frac{H M_P}{M^2} \, ,
\ea
in which $\epsilon = -\dot H/H^2$ is the slow-roll parameter. For the effective field theory description of inflation to be consistent we require $H < M$. On the other hand, the cut-off $M$ can be much smaller than $M_P$. Therefore, one can easily obtain $\beta_0 >1$ while all physical constraints from the non-BD effects are met. In particular, from Eq. (\ref{beta0}) one obtains the upper bound $\beta_0 <  \frac{\sqrt{\epsilon} M_P}{M}$. As a an example, if we take $M \sim 10^{-6} M_P, H \sim 10^{-8} M_P$ and $\epsilon \sim 10^{-2}$, then $\beta_0 \sim 10^3$.  

Although a large value of $\beta_0$ is allowed, one should also take into account the limit imposed from the amplitude of non-Gaussianity $f_{NL}$. As discussed in the Introduction 
a non-BD initial condition can generate large local-type non-Gaussianity. There is tight constraint from PLANCK observation on local-type non-Gaussianity,  $f_{NL}= 2.7 \pm 5.8$
(68 \% CL) \cite{Ade:2013ydc}. The effects of non-BD initial conditions on local-type non-Gaussianity for the models satisfying the ansatz  Eq. (\ref{beta0}) are  studied in  \cite{Ganc:2011dy}. It is shown that $f_{NL}$ is mainly controlled by the value of $\beta_0$ and the relative phase $\varphi_k$ defined in Eq. (\ref{varphi}) (here $\varphi$ and $\beta_0$ are defined for the inflaton perturbations). If one consider 
$\varphi_k \sim k \eta_0 $ then $f_{NL}$ is insensitive to the value of $\beta_0$ so $f_{NL}$ at the order of few can be obtained with no restrictions on the value of $\beta_0$. On the other hand, if one allows $\varphi_k \sim \mathrm{const.}$ then $f_{NL}$ scales like 
  $f_{NL} \sim  \beta_0^4 $. Therefore, $\beta_0$ can not be too large in this limit. 

\begin{figure}
\label{vertex-fig}
\includegraphics[ width=0.5\linewidth]{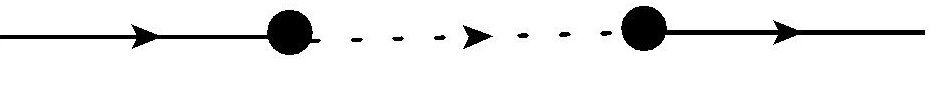}
\caption{ The transfer vertex representing  the interaction of the inflaton field and  the gauge field excitations. The solid line (dashed line) represents the inflaton (the gauge field) propagator and the filled circle represents the coupling factor determined by the interaction Hamiltonian $H_I$ in Eq. (\ref{HI}).  }
\label{Feynman}
\end{figure}

To calculate the anisotropies in power spectrum and bispectrum we use the standard in-in formalism \cite{Weinberg:2005vy, Chen:2010xka, Chen:2009zp} 
\begin{align}
\label{in-in}
\langle Q(\eta_{e})\rangle  = \bigg{\langle} \bigg{[} \overline{T} \exp{\left(i\int_{\eta_{0}}^{\eta_{e}}H_{I}(\eta')d\eta'\right)}\bigg{]} Q(\eta_e)\bigg{[} T \exp{\left(-i\int_{\eta_{0}}^{\eta_{e}}H_{I}(\eta')d\eta'\right)}\bigg{]} \bigg{\rangle} \, ,
\end{align}
in which $Q$ is the physical quantity whose  expectation value is calculated at the end of inflation, $\eta = \eta_e$. Note that for power spectrum, we set $Q= \delta \phi^2$ while for the bispectrum $Q= \delta \phi^3$. In addition,  $T$ and $\overline{T}$ respectively stand for the time-ordered and anti-time-ordered products and $H_{I}$ is the interaction part of the Hamiltonian in the interaction picture.
As for $\eta_0$ we can take $k \eta_0 \rightarrow -\infty$ so the modes of interests were originally deep inside the horizon.

\subsection{Anisotropic  Power Spectrum}

Here we calculate the effects of non-BD initial conditions in anisotropic power spectrum. The corresponding analysis for the BD vacuum are studied in  \cite{Dulaney:2010sq, Gumrukcuoglu:2010yc, Watanabe:2010fh, Bartolo:2012sd,  Funakoshi:2012ym, Yamamoto:2012sq, Emami:2013bk, 
Shiraishi:2013vja}.  

To leading order the anisotropy in inflaton power spectrum, $\delta <{\delta \phi}^2(\eta_{e})>$, is \cite{Emami:2013bk}
\begin{align}
\label{commutator}
\delta <{\delta \phi}^2(\eta_{e})> = - \int_{\eta_{0}}^{\eta_{e}} d\eta_{1} \int_{\eta_{0}}^{\eta_{1}}d\eta_{2} \bigg{[} H_{I}(\eta_{2}) , \bigg{[} H_{I}(\eta_{1}) , {\delta \phi}^2(\eta_e)\bigg{]}\bigg{]} \, .
\end{align}
The second order interaction Hamiltonian, $H_I^{(2)}$, from the Lagrangian Eq. (\ref{L-rho-D1}) is
\ba
\label{HI}
H_I^{(2)} = - \frac{\sqrt{6 I}}{\eta} \frac{b^2 f}{a} \sin^2 \theta  (D_1'^* \delta \phi + D_1' \delta \phi^*) \, .
\ea
The corresponding Feynman diagram is presented in Fig. \ref{Feynman}
which has the form of an exchange vertex linking $\delta \phi$ and $D_1$ in the interaction Hamiltonian Eq. (\ref{HI}).

The fractional change in power spectrum, which is a measure of the anisotropy in power spectrum,  is \cite{Emami:2013bk}
\ba 
\label{powerspec}
\frac{  \delta  \langle \delta \phi(\eta_e)^2  \rangle }{\langle  \delta \phi(\eta_e)^2 \rangle }
=\frac{192 I \sin^4 \theta }{|u(\eta_e)|^2} \int_{\eta_0}^{\eta_e} d \eta_1 \int_{\eta_0}^{\eta_1} d \eta_2
\frac{\eta_1 \eta_2}{\eta_e^4}  Im \left[  u(\eta_1) u(\eta_e)^* \right]
Im \left[  u(\eta_2) u(\eta_e)^*  v'^*(\eta_1) v'(\eta_2) \, .
\right]
\ea
 Furthermore, note that at the end of inflation
$k \eta_e \simeq 0$ so
\ba
u_k(\eta_e) \simeq \frac{i}{\sqrt{2 k} k \eta_e} (\beta^{( \phi)} -\alpha^{( \phi)}) \, .
\ea
In models with BD initial condition one can show that the dominant contributions in integrals in  Eq. (\ref{powerspec}) comes from the super-horizon scales in which $ -1< k\eta_2 \leq k \eta_1 <0$. This is because the contributions of the modes from deep inside the horizon, corresponding to $k\eta \ll -1$, are highly oscillatory so their overall contributions cancel out. However, in the model with non-BD initial conditions the situation is non-trivial. Indeed, it looks challenging how one may calculate the integrals in Eq. (\ref{powerspec})  with
general $\alpha^I$ and $\beta^I$. There are many new terms which did not exist in BD case. The clue, as before, is that the contributions for the modes deep inside the horizon are highly oscillatory. Taking $k \eta_0 \ll -1$ then all the UV oscillations are expected to cancel each other again. As a result, the dominant contributions in the integrals in Eq. (\ref{powerspec}) are expected to come from the super-horizon limit in which 
$k \eta \rightarrow 0^-$. We have checked the validity of this prescription numerically as we explain below.

\begin{figure}
\includegraphics[width=0.9\linewidth]{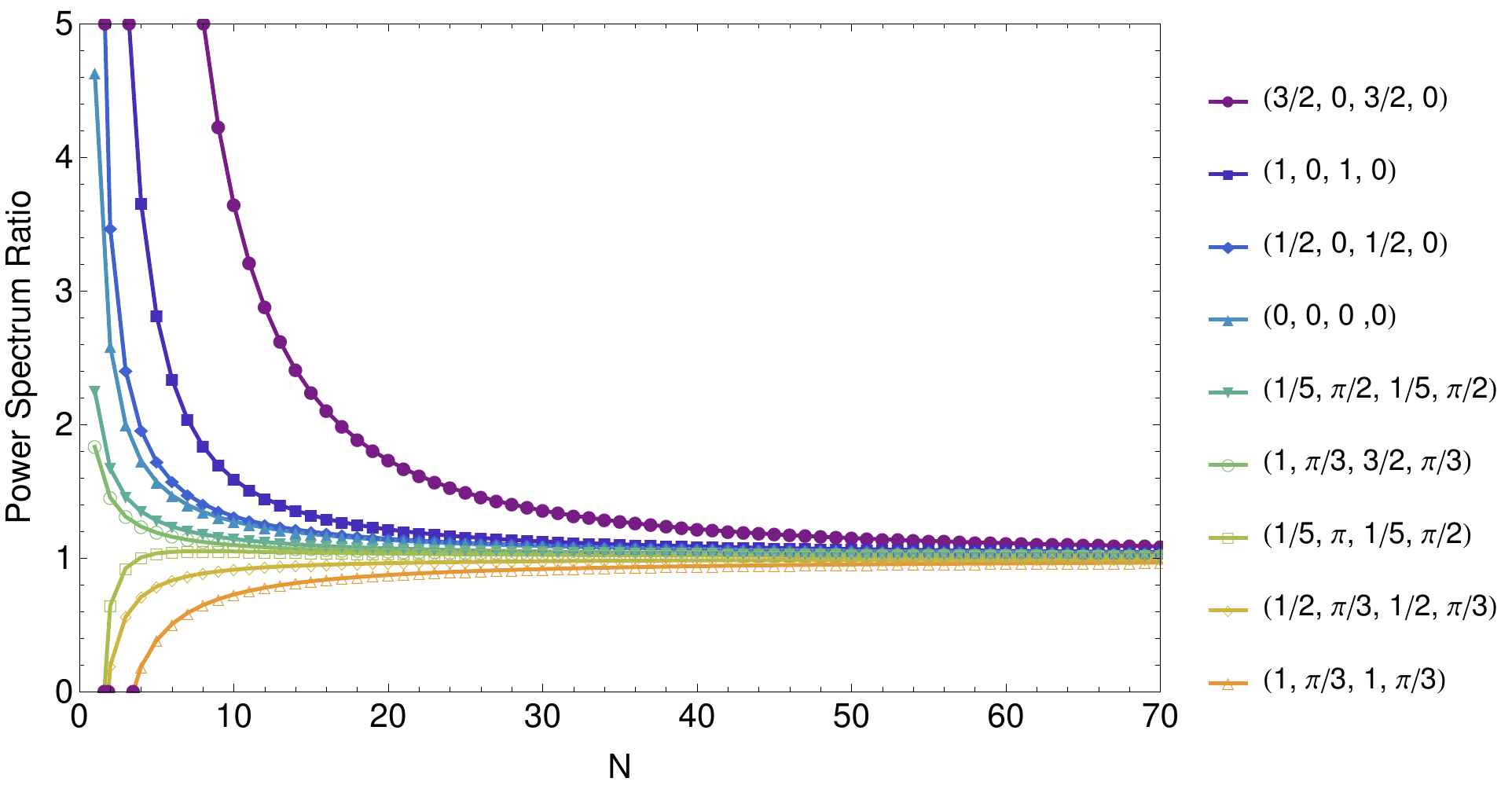}
\caption{ Here the fraction of full numerical calculation Eq. (\ref{powerspec}) to the analytical result Eq. (\ref{analyticpower}) are plotted with  $k\eta_0=  -100$. The four component vectors are defined as ($\chi^{( \phi)}$, $\varphi^{( \phi)}$; $\chi^{( A )}$, $\varphi^{( A )}$). As can be seen this ratio converges rapidly to unity for large $N$. This is
because  the rapid oscillations for the modes deep inside the horizon cancel out and the main
contributions in the in-in integrals comes from the IR region. 
}
\label{power1}
\end{figure}

Taking the super-horizon limit of the integrals in  Eq. (\ref{powerspec}) and neglecting the sub-leading terms we obtain 
\ba
\label{analyticpower}
\frac{  \delta  \langle \delta \phi(\eta_e)^2  \rangle }{\langle  \delta \phi(\eta_e)^2 \rangle }&=&
24I \sin^{2}\theta \, N_e^2 
 \frac{ | \alpha_k^{( A)} - \beta_k^{( A)} |^2   }{  | \alpha_k^{ (\phi) } - \beta_k^{(\phi)} |^2   } 
\nonumber\\
 &=&
 24I  \sin^{2}\theta N_e^2 
 \frac{  \left(1-2\cosh \chi_k^{  ( A  ) }\sinh\chi_k^{  ( A  ) }\cos\varphi_k^{( A )}+2\sinh^{2}\chi_k^{  ( A  ) } \right) }{\left(1-2\cosh \chi_k^{( \phi)}
 \sinh \chi_k^{( \phi)}    \cos\varphi_k^{( \phi)}+2\sinh^{2}\chi_k^{( \phi)} \right)} \, ,
\ea
where $N_e=-\ln (-k \eta_e)$  is the total number of e-foldings counted from the time of end of inflation  and the phase $\varphi_k^I$ is defined as in Eq. (\ref{varphi}).  Note that there are sub-leading terms at the orders ${\cal O}(N_e)$ and ${\cal O}(1)$ which we have neglected.
More specifically, the $N_e$ dependence above has the sub-leading corrections 
$N_e^2 \rightarrow N_e^2 \left(  1+ \frac{1}{6 N_e} \right) + {\cal O}(1)$. In the limit of our interest in  which $N_e \gg 1$ we can neglect the  order ${\cal O}(N_e)$ and ${\cal O}(1)$ corrections. 

Correspondingly, the anisotropy parameter $g_*$ defined in Eq. (\ref{g-def}) is obtained to be 
\ba
\label{g-nBD}
g_* &=& -24I N_e^2 
\frac{ | \alpha_k^{( A)} - \beta_k^{( A)} |^2   }{  | \alpha_k^{ (\phi) } - \beta_k^{(\phi)} |^2   }  \nonumber\\
&=& -24I N_e^2 
\frac{  \left(1-2\cosh \chi_k^{  ( A  ) }\sinh\chi_k^{  ( A  ) }\cos\varphi_k^{( A )}+2\sinh^{2}\chi_k^{  ( A  ) } \right) }{\left(1-2\cosh \chi_k^{( \phi)}
 \sinh \chi_k^{( \phi)}    \cos\varphi_k^{( \phi)}+2\sinh^{2}\chi_k^{( \phi)} \right)} \, .
\ea
As expected, in the BD limit in which $\alpha_k^I =1, \beta_k^I=0$ 
we get the known result  $g_* = -24 I N_e^2$. 

One simple observation from the above formula is that $g_* <0$ so the inclusion of the non-BD initial condition does not change the sign of the anisotropy parameter. The constraints from the PLANCK data implies $| g_*| \lesssim 10^{-2}$ \cite{Kim:2013gka}. In models with the BD initial condition this imposes the strong fine-tuning $I < 10^{-6}$. However, in the model with non-BD initial conditions, we have enough new parameter to bypass this fine-tuning. There are different options to enhance the factor $  | \alpha_k^{( A)} - \beta_k^{( A)} |^2  /  | \alpha_k^{ (\phi) } - \beta_k^{(\phi)} |^2  $.  For example, let us take the inflaton field perturbations to be in BD vacuum so $\alpha_k^{( \phi)}=1,  \beta_k^{( \phi)}=0$.
If we take $\cos \varphi_k^{( A )} \sim -1$ and $\chi_k^{  ( A  ) } \sim 4$, then the fraction in 
Eq. (\ref{g-nBD}) can be as large as $10^3$. This can help to relax the fine-tuning on the anisotropy parameter and  one can easily satisfy the  observational constraints on $g_*$ with $I\sim 10^{-3}$.  For large values of $\chi_k$ the role of the phase $\varphi_k$ is important. For example, taking the BD initial condition for the scalar field perturbation and assuming $\varphi_k^{( A )}\sim \pi$ and $\chi_k^{  ( A  ) }\gg 1$  then $g_*$ is enhanced by  the factor  $e^{\chi_k^{  ( A  ) }}$. Having this said, one can not take $\chi_k$ arbitrary large because one has to take into account the back-reaction effects and the constraints from the amplitude of local non-Gaussianity as discussed after Eq. (\ref{beta0}).  Finally in the limit that $\varphi_k^{( A )}=\varphi_k^{( \phi)}$ and $\chi_k^{  ( A  ) }=\chi_k^{( \phi)}$ the effect of non-BD correction is canceled from \eqref{g-nBD}. One can also show that for $\varphi_k^{( A )},\,\,\varphi_k^{( \phi)}\sim 0$ and $\chi_k^{  ( A  ) },\,\,\chi_k^{( \phi)}\gg 1$ the same conclusion is obtained.

\begin{figure}
\begin{center}
\includegraphics[width=1\linewidth]{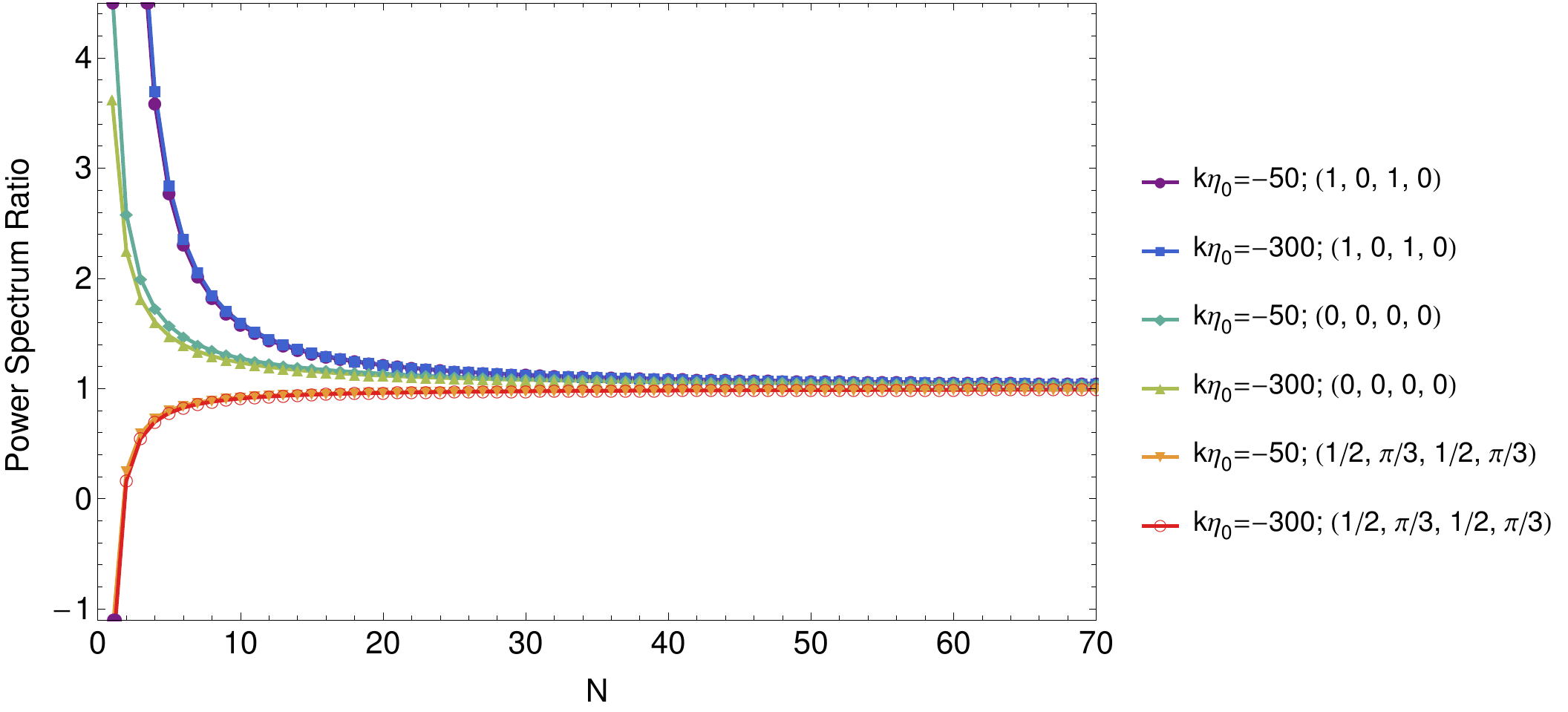}
\caption{ The same plot as in Fig. \ref{power1}. For a given value of the  parameters ($\chi^{( \phi)}$, $\varphi^{( \phi)}$; $\chi^{( A )}$, $\varphi^{( A )}$) there are two curves with $k \eta_0=-50 $ and $k \eta_0=-300$. For each set of ($\chi^{( \phi)}$, $\varphi^{( \phi)}$; $\chi^{( A )}$, $\varphi^{( A )}$) the two curves coincide very well with each other. This indicates that the results of in-in integrals are insensitive to $k \eta_0$ as long as $k\eta_0 \ll -1$. 
\label{power2}}
\end{center}
\end{figure}

In Fig. \ref{power1} we have plotted the ratio of the in-in integrals in Eq. (\ref{powerspec})
obtained full-numerically to the analytical result Eq. (\ref{analyticpower}) with $k \eta_0 =-100$.  As can be seen, for large $N_e$, the ratio approaches unity rapidly. This justifies our analytical  methods of calculating the in-in integrals in which the super-horizon contributions of the integrals are kept while the contributions from the rapid oscillations deep in the UV regions are discarded. The deviation from unity for small $N_e$ is due to sub-leading 
corrections.  As expected, for 
large $N_e$ the approximation becomes more and more accurate. In Fig. \ref{power2} we have plotted this ratio for different values of $k \eta_0$. As expected, as long as $k \eta_0 \ll -1$, the results are independent of $\eta_0$. This is again a manifestation of the fact that the dominant contributions to the in-in integrals are from super-horizon regions in which $-1 \leq k \eta <0$ and the UV contributions in the regions $k \eta_0 \leq k\eta \leq-1$ cancel out.

\subsection{Anisotropic Bispectrum}

Here we calculate the the effects of non-BD initial conditions on anisotropic bispectrum. The anisotropic bispectrum for models of anisotropic inflation and related models were calculated in  \cite{ Bartolo:2012sd,  Funakoshi:2012ym, Yamamoto:2012sq, 
Shiraishi:2013vja, Abolhasani:2013zya,  Abolhasani:2013bpa, Lyth:2013sha, Jain:2012vm, 
Nurmi:2013gpa, Fujita:2013qxa, Shiraishi:2013sv, Baghram:2013lxa, Thorsrud:2013kya, Rodriguez:2013cj, BeltranAlmeida:2011db}, for a recent work on
trispectrum see \cite{Shiraishi:2013oqa}. As  proved in \cite{Emami:2013bk}, the leading contribution in the bispectrum comes from the gauge fields perturbations and one can neglect the non-Gaussianities generated from the metric excitations. The contributions of the metric fluctuations in bispectrum are at the order of the slow roll parameter following Maldacena's analysis \cite{Maldacena:2002vr}. 

The cubic interaction $H_I^{(3)}$  between the inflaton and the gauge field comes from the
gauge kinetic coupling term $f(\phi)^2 F_{\mu \nu} F^{\mu \nu}$ which has the following form
\ba
H^{(3)}_{I} =  - f(\phi)f_{,\phi}(\phi) \int d^{3}p \int d^{3}k  \sin{\theta_{p}} \sin{\theta_{k}} \cos{\left(\theta_{k} - \theta_{p}\right)} \delta \phi_{-\left(p+k\right)} D_{1k}' D_{1p}'
\ea
in which  $\theta_k$ is  the angle between the momentum direction $\bk$
and the $x$-axis as defined in Eq. (\ref{ab}). The form of this cubic Hamiltonian is intuitively understandable. Because of the attractor solution, the gauge field at the background leads to the factor $\sqrt{R}$ so in order to get the leading interaction we should consider the second order terms for the gauge field fluctuations and the first order term for the inflaton field.

Now by using the transfer vertex, Eq. (\ref{HI}), we can calculate the anisotropic 
bispectrum. Following the general prescription of the in-in formalism given in Eq. (\ref{in-in})
the leading contribution in the bispectrum comes from the third ordered expansion as follows
\ba
 \langle \calR^{3} (\eta_e) \rangle = -\frac{i}{\left(2 \epsilon \right)^{3/2}M_{P}^3}\int_{\eta_0}^{\eta_e} d \eta_1 \int_{\eta_0}^{\eta_1} d \eta_2
\int_{\eta_0}^{\eta_2} d \eta_3 \bigg{[} H_{I}(\eta_{3}), \bigg{[} H_{I}(\eta_{2}),\bigg{[} H_{I}(\eta_{1}), \delta \phi^{3}(\eta_e) \bigg{]} \bigg{]} \bigg{]}
\ea
in which the relation $\calR = \delta \phi/\sqrt{2 \epsilon} M_P$ have been used for the comoving curvature perturbation $\calR$. 

Even in the absence of anisotropies large bispectrum can be generated with the non-BD initial conditions  \cite{Agullo:2010ws, Ganc:2011dy, Chialva:2011hc, Creminelli:2011rh,
Ganc:2012ae, Agullo:2012cs, Agarwal:2012mq, Ashoorioon:2010xg, Chen:2006nt,Holman:2007na, Meerburg:2009ys, Ashoorioon:2013eia, Kundu:2013gha, Brahma:2013rua, 
Bahrami:2013isa, Gong:2013yvl}. However, we are not interested in isotropic bispectrum so in the following analysis we concentrate only on anisotropic bispectrum. 

To calculate the anisotropic bispectrum from the above nested integrals we should replace one of $H_{I}(\eta_i)$ with  the cubic Hamiltonian $H^{(3)}_{I}$ and the rest with the quadratic $H^{(2)}_{I}$. Since there are three different locations for  $H^{(3)}_{I}$ to sit down, we will have three possible terms. However, as we will see the leading result is proportional to $N_e^{3}$ which means that these different situations are equivalent. So  in the following we only mention the final answer which has been multiplied by the  factor 3,
\ba
\label{bi-ani}
 \langle \calR_{\bk_1}(\eta_e)  \calR_{\bk_2} (\eta_e) \calR_{\bk_3} (\eta_e) \rangle
 {\big |_{\mathrm{aniso.}}}
&=& - \frac{2^5 \times 72 I}{ (2 \epsilon)^{3/2} M_P^3}
(2 \pi)^3 \delta^3(\bk_1 + \bk_2 + \bk_3)
\left\{
 \sin^3 \theta_{k2}  \sin^3 \theta_{k3}  \cos (\theta_{k2} -  \theta_{k3} )  \right.  \\
&\times&\left. \int_{\eta_0}^{\eta_e} d \eta_1 \int_{\eta_0}^{\eta_1} d \eta_2
\int_{\eta_0}^{\eta_2} d \eta_3
\frac{f(\eta_3)}{\eta_3}  \frac{f(\eta_2)}{\eta_2}
\frac{f(\eta_1) f_{, \phi} (\eta_1)}{a(\eta_1)} {\rm Im} \left[ u_{k_1}(\eta_1) u_{k_1}(\eta_e)^*  \right]  \right.\nonumber\\
&\times &  \left.
{\rm Im} \left[  u_{k_2}(\eta_2) u_{k_2}(\eta_e)^*  v_{k_2}'^*(\eta_1) v_{k_2}'(\eta_2) \right]
{\rm Im} \left[  u_{k_3}(\eta_3) u_{k_3}(\eta_e)^*  v_{k_3}'^*(\eta_1) v_{k_3}'(\eta_3) \right]
 + 2 \, \mathrm{c.p.} \right\}   \nonumber
\ea
in which c.p. represents the cyclic permutations so we have three terms in total. 

Let us define
\ba
x_0 = k_3 \eta_0  \quad , \quad
x_1 = k_1 \eta_1  \quad , \quad  x_2 = k_2 \eta_2  \quad , \quad  x_3= k_3 \eta_3\quad , \quad
x_e = k_1 \eta_e \, .
\ea 
In the integral  above, $k_1$ appears differently than $k_2 $ and $k_3$ so that is why we have defined $x_e$ with respect to $k_1$. 

Noting that $\phi= \sqrt{2} M_P/\sqrt{\epsilon}$ and the isotropic power spectrum  is given by
\ba
P_\calR(k) = \frac{H^2}{4 k^3 \epsilon_H M_P^2}   | \alpha_k^{( \phi)} - \beta_k^{( \phi)} |^2\, ,
\ea
Eq. (\ref{bi-ani}) yields
\ba
 \langle \calR_{\bk_1}  \calR_{\bk_2} \calR_{\bk_3}  \rangle {\big |_{\mathrm{aniso.}}}
 &=& ( -72 \times 2^8 \times I )  (2 \pi)^3 \delta^3(\bk_1 + \bk_2 + \bk_3)
 \left\{  \frac{k_2 k_3}{k_1}   P_\calR(\bk_2) P_\calR(\bk_3) C(\bk_2, \bk_3) \right. \nonumber\\
&& | \alpha_{k_2}^{( \phi)} - \beta_{k_2}^{( \phi)} |^{-2}  | \alpha_{k_3}^{( \phi)} - \beta_{k_3}^{( \phi)} |^{-2}
\left. \int_{x_0}^{x_e} dx_1 \int_{x_0}^{x_1} dx_2 \int_{x_0}^{x_2} dx_3  \frac{x_1^5 x_2 x_3}{x_e^5} {\rm Im} \left[ u_{k_1}(\eta_1) u_{k_1}(\eta_e)^*  \right]  \right. \nonumber\\
&&\left.
 {\rm Im} \left[  u_{k_2}(\eta_2) u_{k_2}(\eta_e)^*  v_{k_2}'^*(\eta_1) v_{k_2}'(\eta_2) \right] {\rm Im} \left[  u_{k_3}(\eta_3) u_{k_3}(\eta_e)^*  v_{k_3}'^*(\eta_1) v_{k_3}'(\eta_3) \right] + 2 \mathrm{c.p.}  \right\}   \, ,
\ea
in which the shape function $C(\bk_2, \bk_3)$ is defined as \cite{ Bartolo:2012sd, Abolhasani:2013zya}
\ba
C(\bk_2, \bk_3) &\equiv& 1- (\widehat \bn. \widehat \bk_2)^2 -  (\widehat \bn. \widehat \bk_3)^2 + 
(\widehat \bn. \widehat \bk_2) (\widehat \bn. \widehat \bk_3) (\widehat \bk_2 . \widehat \bk_3)
\nonumber\\
&=&  \sin \theta_{k2}  \sin \theta_{k3}  \cos (\theta_{k2} -  \theta_{k3} )  \, .
\ea

We are interested in the bispectrum defined as
\ba
\label{bi- def}
\langle  \calR_{\bk_1}  \calR_{\bk_2}  \calR_{\bk_3} \rangle &\equiv& \left( 2 \pi \right)^3 \delta^3 \left( \bk_{1} +  \bk_{2} +  \bk_{3}\right) B_{\calR}(\bk_{1}, \bk_{2}, \bk_{3}) \, .
\ea
Therefore, the anisotropic bispectrum is obtained to be
\ba 
\label{numbispectrum}
B_{\calR}^{\mathrm{aniso.}}(\bk_{1}, \bk_{2}, \bk_{3})  &=&   (- 72 \times 2^8 \times I )
\left\{   \frac{k_2 k_3}{k_1} P_\calR(\bk_2) P_\calR(\bk_3) C(\bk_2, \bk_3)
 \int_{x_0}^{x_e} dx_1 \int_{x_0}^{x_1} dx_2 \int_{x_0}^{x_2} dx_3  \frac{x_1^5 x_2 x_3}{x_e^5} \right. \nonumber \\
&& | \alpha_{k_2}^{( \phi)} - \beta_{k_2}^{( \phi)} |^{-2}  | \alpha_{k_3}^{( \phi)} - \beta_{k_3}^{( \phi)} |^{-2}
\times {\rm Im}   \left[ u_{k_1}(\eta_1) u_{k_1}(\eta_e)^*  \right]
\times  {\rm Im} \left[  u_{k_2}(\eta_2) u_{k_2}(\eta_e)^*  v_{k_2}'^*(\eta_1) v_{k_2}'(\eta_2) \right]  \nonumber \\
&&\left.
\times {\rm Im} \left[  u_{k_3}(\eta_3) u_{k_3}(\eta_e)^*  v_{k_3}'^*(\eta_1) v_{k_3}'(\eta_3) \right]
+ 2 \mathrm{c.p.}  \right\} 
\ea
The above nested integral has very complicated forms in the presence of non-BD initial conditions. It seems impossible to evaluate this integral in general case. However, as we discussed in the power spectrum case, we are interested in the limit where the modes are 
 initially deep inside the horizon and  $k\eta_0 \ll -1$ so $x_0 \ll -1$. As a result  the contributions of these UV modes in the in-in integrals cancel out because of the rapid oscillations. Therefore, one has to consider the contributions of the modes after the time of horizon crossing. This correspond to taking the integral as $  \int_{-1}^{x_e}   \int_{-1}^{k_2 x_1/k_1} \int_{-1}^{k_3 x_2/k_2}...$  with ... representing the integrand function. Similar to the power spectrum case we have checked that our analytical results obtained this way converges to the full numerical results
for $N_e \gg 1$.  
 
With this prescription, the bispectrum is calculated to be 
\ba
\label{fnl leading}
B_\calR^{\mathrm{aniso.}}( \bk_1,  \bk_2,  \bk_3)=  288 I N(k_1) N(k_2) N(k_3) 
\left[ \frac{ | \alpha_{k_2}^{( A )} - \beta_{k_2}^{( A )} |^{2}  | \alpha_{k_3}^{( A )} - \beta_{k_3}^{( A )} |^{2}}{ | \alpha_{k_2}^{( \phi)} - \beta_{k_2}^{( \phi)} |^{2}  | \alpha_{k_3}^{( \phi)} - \beta_{k_3}^{( \phi)} |^{2}}
C(\bk_1, \bk_2) P_\calR(k_{1})P_\calR(k_{2}) \, 
+ 2 \mathrm{c.p.} \right]  
\ea
Here $N(k_i)$ represents the number of e-folds when the  mode $\bk_i$ has left the horizon.
For practical purpose one may simply take $N(k_i) \simeq N_e$.  In the limit of BD vacuum, the above formula reproduces the results in \cite{ Bartolo:2012sd, Abolhasani:2013zya}. 

It is also instructive to calculate $f_{NL}$ in the squeezed limit $k_1 \ll k_2 \simeq k_3$ defined via
\ba
f_{NL} (\bk_1, \bk_2, \bk_3) = \lim_{k_1 \rightarrow 0} \frac{5}{12}
\frac{B_\calR(\bk_1, \bk_2, \bk_3)}{P_\calR(k_1) P_{\calR}(k_2)} \, .
\ea
In this limit, for the anisotropic part of $f_{NL}$, we get 
\ba
\label{fnl squeezed}
f_{NL}^{\mathrm{aniso.}} = 240 I N(k_{1}) N(k_{2})^2 C(\bk_{1}, \bk_{2})    
\frac{ | \alpha_{k_2}^{( A )} - \beta_{k_2}^{( A )} |^{4} }{ | \alpha_{k_2}^{( \phi)} - \beta_{k_2}^{( \phi)} |^{4}}
\quad \quad  \quad
 ( k_{1} \ll k_{2} \simeq k_{3} )  
\ea
It will be helpful to eliminate the unknown factors $| \alpha_{k_2}^{I} - \beta_{k_2}^{I} |$
and express $f_{NL}^{\mathrm{aniso.}}$ in terms of the observational parameter $g_*$ obtained in  Eq. (\ref{g-nBD}). This yields 
\ba
f_{NL}^{\mathrm{aniso.}} = \frac{10}{24} \frac{g_*^2 N(k_1)}{I N(k_2)^2} \sim \frac{g_*^2}{100 I} 
\quad \quad  \quad
 ( k_{1} \ll k_{2} \simeq k_{3} ) 
\ea
This expression indicates that for a fixed value of $g_*$, the larger the value of $I$ the smaller the value of $f_{NL}^{\mathrm{aniso.}}$. However, to satisfy the observational constraints from PLANCK we need $|g_*| \lesssim 10^{-2}$ \cite{Kim:2013gka}.  In the BD case with $N_e =60$ this yields  $I \sim 10^{-6}$ and one can obtain $f_{NL}^{\mathrm{aniso.}}  \sim \mathrm{few}$. However, in the presence of non-BD initial condition, if we use the additional freedom in Eq. (\ref{g-nBD}) to lower the fine-tuning on $I$, say $I$  as large as $I \sim 10^{-3}$, then $f_{NL}^{\mathrm{aniso.}}$ becomes very small, comparable to slow-roll parameters. Of course, in this limit we can not neglect the gravitational back-reactions \cite{Maldacena:2002vr}. In addition in this limit the anisotropic bispectrum is too small to be detected. \\

To summarize, in this work the effects of a generalized non-vacuum initial state on anisotropic power spectrum and bispectrum were studied. The motivation originates from the fact that in these models the level of anisotropies grow with the total number of e-folds ($N_e^2$ for power spectrum and $N_e^3$ for the bispectrum). This is because the gauge field fluctuations accumulate on super-horizon scales to make the classical back-ground more and more anisotropic. As a result, the total number of e-foldings in models of anisotropic inflation can not be arbitrarily large. As a result, the physical predictions are sensitive to the initial state at the start of inflation. Intuitively speaking, since inflation can not have an extended period in the past in these models, then there are remnants of initial quantum state of the universe which 
did not settle down to vacuum. We have parameterized the deviation from the Bunch-Davies or  Minkowski vacuum state in terms of the Bogoliubov  coefficients $\alpha_k^{I}$ and $\beta_k^I$. As expected, the predictions on anisotropic power spectrum and bispectrum depends on these  non Bunch-Davies coefficients. We also allowed for different Bogoliubov  coefficients for the inflaton and the gauge field fluctuations. This is motivated from the intuition that the gauge field and the inflaton may have been affected differently in the past inflationary history. For example, there may be a phase transition or particle creation in the inflaton sector which might not  affect the gauge field sector. We have shown that in terms of the Bogoliubov  coefficients the physical parameters $g_{*}$ and  $f_{NL}^{\mathrm{aniso.}}$ scales like
$g_* \propto |\frac{  \alpha^{(A)} - \beta^{(A)} }{  \alpha^{( \phi)} - \beta^{( \phi)} }|^2$ and $f_{NL}^{\mathrm{aniso.}} \propto |\frac{  \alpha^{(A)} - \beta^{(A)}  }{  \alpha^{( \phi)} - \beta^{( \phi)} }|^4$. It is argued that one may use these additional factors to reduce the level of fine-tuning on the anisotropy parameter $I$ while satisfying the observational constraints on $g_*$. In addition, for a fixed value of $g_*$ the value of $f_{NL}$ is decreased by increasing the value of $I$.

\acknowledgments

We would like to thank A. A. Abolhasani and Y. Wang   for useful discussions. R. E. would like to thank ICTP for the hospitality during the course of this work. H.F. would like to thank KITPC for the hospitality during the workshop `` Cosmology after PLANCK''  and ICTP for the hospitality where this work was in progress.




\section*{References}
\begin{thebibliography}{}


\bibitem{Bennett:2012zja}
  C.~L.~Bennett {\it et al.}  [WMAP Collaboration],
  ``Nine-Year Wilkinson Microwave Anisotropy Probe (WMAP) Observations: Final Maps and Results,''
  arXiv:1212.5225 [astro-ph.CO].

\bibitem{Hinshaw:2012fq}
  G.~Hinshaw {\it et al.} [WMAP Collaboration]
  ``Nine-Year Wilkinson Microwave Anisotropy Probe (WMAP) Observations: Cosmological Parameter Results,''
  arXiv:1212.5226 [astro-ph.CO].

\bibitem{Ade:2013lta}
  P.~A.~R.~Ade {\it et al.}  [Planck Collaboration],
  ``Planck 2013 results. XVI. Cosmological parameters,''
  arXiv:1303.5076 [astro-ph.CO].

\bibitem{Ade:2013uln}
  P.~A.~R.~Ade {\it et al.}  [Planck Collaboration],
  ``Planck 2013 results. XXII. Constraints on inflation,''
  arXiv:1303.5082 [astro-ph.CO].

\bibitem{Ade:2013nlj}
  P.~A.~R.~Ade {\it et al.}  [Planck Collaboration],
  ``Planck 2013 results. XXIII. Isotropy and Statistics of the CMB,''
  arXiv:1303.5083 [astro-ph.CO].

\bibitem{Kim:2013gka} 
  J.~Kim and E.~Komatsu,
  ``Limits on anisotropic inflation from the Planck data,''
  Phys.\ Rev.\ D {\bf 88}, 101301 (2013)
  [arXiv:1310.1605 [astro-ph.CO]].


\bibitem{Turner:1987bw}
  M.~S.~Turner and L.~M.~Widrow,
  ``Inflation Produced, Large Scale Magnetic Fields,''
  Phys.\ Rev.\ D {\bf 37}, 2743 (1988).


\bibitem{Ratra:1991bn}
  B.~Ratra,
  ``Cosmological 'seed' magnetic field from inflation,''
  Astrophys.\ J.\  {\bf 391}, L1 (1992).


\bibitem{Soda:2012zm} 
  J.~Soda,
  ``Statistical Anisotropy from Anisotropic Inflation,''
  Class.\ Quant.\ Grav.\  {\bf 29}, 083001 (2012)
  [arXiv:1201.6434 [hep-th]].

\bibitem{Maleknejad:2012fw} 
  A.~Maleknejad, M.~M.~Sheikh-Jabbari and J.~Soda,
  ``Gauge Fields and Inflation,''
  Phys.\ Rept.\  {\bf 528}, 161 (2013)
  [arXiv:1212.2921 [hep-th]].






\bibitem{Watanabe:2009ct}
  M.~a.~Watanabe, S.~Kanno and J.~Soda,
  ``Inflationary Universe with Anisotropic Hair,''
  Phys.\ Rev.\ Lett.\  {\bf 102}, 191302 (2009)
  [arXiv:0902.2833 [hep-th]].
  
\bibitem{Ohashi:2013pca} 
  J.~Ohashi, J.~Soda and S.~Tsujikawa,
  ``Anisotropic power-law k-inflation,''
  Phys.\ Rev.\ D {\bf 88}, 103517 (2013)
  [arXiv:1310.3053 [hep-th]].

\bibitem{Ohashi:2013mka} 
  J.~Ohashi, J.~Soda and S.~Tsujikawa,
  ``Anisotropic Non-Gaussianity from a Two-Form Field,''
  Phys.\ Rev.\ D {\bf 87}, 083520 (2013)
  [arXiv:1303.7340 [astro-ph.CO]].

\bibitem{Emami:2010rm} 
  R.~Emami, H.~Firouzjahi, S.~M.~Sadegh Movahed and M.~Zarei,
  ``Anisotropic Inflation from Charged Scalar Fields,''
  JCAP {\bf 1102}, 005 (2011)
  [arXiv:1010.5495 [astro-ph.CO]].

\bibitem{Thorsrud:2012mu} 
  M.~Thorsrud, D.~F.~Mota and S.~Hervik,
  ``Cosmology of a Scalar Field Coupled to Matter and an Isotropy-Violating Maxwell Field,''
  JHEP {\bf 1210}, 066 (2012)
  [arXiv:1205.6261 [hep-th]].




\bibitem{Dulaney:2010sq}
  T.~R.~Dulaney, M.~I.~Gresham,
  ``Primordial Power Spectra from Anisotropic Inflation,''
  Phys.\ Rev.\  {\bf D81}, 103532 (2010).
  [arXiv:1001.2301 [astro-ph.CO]].


\bibitem{Gumrukcuoglu:2010yc}
  A.~E.~Gumrukcuoglu, B.~Himmetoglu, M.~Peloso,
  ``Scalar-Scalar, Scalar-Tensor, and Tensor-Tensor Correlators from Anisotropic Inflation,''
  Phys.\ Rev.\  {\bf D81}, 063528 (2010).
  [arXiv:1001.4088 [astro-ph.CO]].



\bibitem{Watanabe:2010fh}
  M.~a.~Watanabe, S.~Kanno and J.~Soda,
  ``The Nature of Primordial Fluctuations from Anisotropic Inflation,''
  Prog.\ Theor.\ Phys.\  {\bf 123}, 1041 (2010)
  [arXiv:1003.0056 [astro-ph.CO]].

\bibitem{Bartolo:2012sd} 
  N.~Bartolo, S.~Matarrese, M.~Peloso and A.~Ricciardone,
  ``The anisotropic power spectrum and bispectrum in the $f(phi) F^2$ mechanism,''
  Phys.\ Rev.\ D {\bf 87}, 023504 (2013)
  [arXiv:1210.3257 [astro-ph.CO]].



\bibitem{Funakoshi:2012ym}
  H.~Funakoshi and K.~Yamamoto,
  ``Primordial bispectrum from inflation with background gauge fields,''
  arXiv:1212.2615 [astro-ph.CO].

\bibitem{Yamamoto:2012sq}
  K.~Yamamoto,
  ``Primordial Fluctuations from Inflation with a Triad of Background Gauge Fields,''
  Phys.\ Rev.\ D {\bf 85}, 123504 (2012)
  [arXiv:1203.1071 [astro-ph.CO]].


\bibitem{Emami:2013bk} 
  R.~Emami and H.~Firouzjahi,
  ``Curvature Perturbations in Anisotropic Inflation with Symmetry Breaking,''
  JCAP {\bf 1310}, 041 (2013)
  [arXiv:1301.1219 [hep-th]].


\bibitem{Shiraishi:2013vja} 
  M.~Shiraishi, E.~Komatsu, M.~Peloso and N.~Barnaby,
  ``Signatures of anisotropic sources in the squeezed-limit bispectrum of the cosmic microwave background,''
  JCAP {\bf 1305}, 002 (2013)
  [arXiv:1302.3056 [astro-ph.CO]].


\bibitem{Abolhasani:2013zya} 
  A.~A.~Abolhasani, R.~Emami, J.~T.~Firouzjaee and H.~Firouzjahi,
  ``$\delta N$ formalism in anisotropic inflation and large anisotropic bispectrum and trispectrum,''
  JCAP {\bf 1308}, 016 (2013)
  [arXiv:1302.6986 [astro-ph.CO]].


\bibitem{Abolhasani:2013bpa} 
  A.~A.~Abolhasani, R.~Emami and H.~Firouzjahi,
  ``Primordial Anisotropies in Gauged Hybrid Inflation,''
  arXiv:1311.0493 [hep-th].
  
  
\bibitem{Chen:2013tna} 
  X.~Chen and Y.~Wang,
  ``Non-Bunch-Davies Anisotropy,''
  arXiv:1306.0609 [hep-th].

\bibitem{Chen:2013eaa} 
  X.~Chen and Y.~Wang,
  ``Relic Vector Field and CMB Large Scale Anomalies,''
  arXiv:1305.4794 [astro-ph.CO].
 
  
  
  


\bibitem{Agullo:2010ws}
  I.~Agullo and L.~Parker,
  ``Non-gaussianities and the Stimulated creation of quanta in the inflationary universe,''
  Phys.\ Rev.\ D {\bf 83}, 063526 (2011).
  [arXiv:1010.5766 [astro-ph.CO]].

\bibitem{Ganc:2011dy}
  J.~Ganc,
  ``Calculating the local-type fNL for slow-roll inflation with a non-vacuum initial state,''
  Phys.\ Rev.\ D {\bf 84}, 063514 (2011).
  [arXiv:1104.0244 [astro-ph.CO]].

\bibitem{Chialva:2011hc}
  D.~Chialva,
  ``Signatures of very high energy physics in the squeezed limit of the bispectrum (violation of Maldacena's condition),''
  JCAP {\bf 1210}, 037 (2012).
  [arXiv:1108.4203 [astro-ph.CO]].

\bibitem{Creminelli:2011rh}
  P.~Creminelli, G.~D'Amico, M.~Musso and J.~Norena,
  ``The (not so) squeezed limit of the primordial 3-point function,''
  JCAP {\bf 1111}, 038 (2011)
  [arXiv:1106.1462 [astro-ph.CO]].

\bibitem{Ganc:2012ae}
  J.~Ganc and E.~Komatsu,
  ``Scale-dependent bias of galaxies and mu-type distortion of the cosmic microwave background spectrum from single-field inflation with a modified initial state,''
  Phys.\ Rev.\ D {\bf 86}, 023518 (2012).
  [arXiv:1204.4241 [astro-ph.CO]].

\bibitem{Agullo:2012cs}
  I.~Agullo and S.~Shandera,
  ``Large non-Gaussian Halo Bias from Single Field Inflation,''
  JCAP {\bf 1209}, 007 (2012).
  [arXiv:1204.4409 [astro-ph.CO]].

\bibitem{Agarwal:2012mq}
  N.~Agarwal, R.~Holman, A.~J.~Tolley and J.~Lin,
  ``Effective field theory and non-Gaussianity from general inflationary states,''
  JHEP {\bf 1305}, 085 (2013)
  [arXiv:1212.1172 [hep-th]].

\bibitem{Ashoorioon:2010xg}
  A.~Ashoorioon and G.~Shiu,
  ``A Note on Calm Excited States of Inflation,''
  JCAP {\bf 1103}, 025 (2011)
  [arXiv:1012.3392 [astro-ph.CO]].

\bibitem{Chen:2006nt}
  X.~Chen, M.~-x.~Huang, S.~Kachru and G.~Shiu,
  ``Observational signatures and non-Gaussianities of general single field inflation,''
  JCAP {\bf 0701}, 002 (2007).

\bibitem{Holman:2007na}
  R.~Holman and A.~J.~Tolley,
  ``Enhanced Non-Gaussianity from Excited Initial States,''
  JCAP {\bf 0805}, 001 (2008).

\bibitem{Meerburg:2009ys}
  P.~D.~Meerburg, J.~P.~van der Schaar and P.~S.~Corasaniti,
  ``Signatures of Initial State Modifications on Bispectrum Statistics,''
  JCAP {\bf 0905}, 018 (2009).


\bibitem{Ashoorioon:2013eia} 
  A.~Ashoorioon, K.~Dimopoulos, M.~M.~Sheikh-Jabbari and G.~Shiu,
  ``Reconciliation of High Energy Scale Models of Inflation with Planck,''
  arXiv:1306.4914 [hep-th].

\bibitem{Kundu:2013gha} 
  S.~Kundu,
  ``Non-Gaussianity Consistency Relations, Initial States and Back-reaction,''
  arXiv:1311.1575 [astro-ph.CO].

\bibitem{Brahma:2013rua} 
  S.~Brahma, E.~Nelson and S.~Shandera,
  ``Fossilized Relics and Primordial Clocks,''
  arXiv:1310.0471 [astro-ph.CO].

\bibitem{Bahrami:2013isa} 
  S.~Bahrami and ƒa.~ƒ.Flanagan,
  ``Primordial non-Gaussianities in single field inflationary models with non-trivial initial states,''
  arXiv:1310.4482 [astro-ph.CO].
  
\bibitem{Gong:2013yvl} 
  J.~-O.~Gong and M.~Sasaki,
  ``Squeezed primordial bispectrum from general vacuum state,''
  Class.\ Quant.\ Grav.\  {\bf 30}, 095005 (2013)
  [arXiv:1302.1271 [astro-ph.CO]].

\bibitem{Maldacena:2002vr} 
  J.~M.~Maldacena,
  ``Non-Gaussian features of primordial fluctuations in single field inflationary models,''
  JHEP {\bf 0305}, 013 (2003)
  [astro-ph/0210603].
  
  
  
  
  

\bibitem{Namjoo:2012aa}
  M.~H.~Namjoo, H.~Firouzjahi and M.~Sasaki,
  ``Violation of non-Gaussianity consistency relation in a single field inflationary model,''
  Europhys.\ Lett.\  {\bf 101}, 39001 (2013)
  [arXiv:1210.3692 [astro-ph.CO]].

\bibitem{Chen:2013aj}
  X.~Chen, H.~Firouzjahi, M.~H.~Namjoo and M.~Sasaki,
  ``A Single Field Inflation Model with Large Local Non-Gaussianity,''
  Europhys.\ Lett.\  {\bf 102}, 59001 (2013)
  [arXiv:1301.5699 [hep-th]].

\bibitem{Huang:2013lda}
  Q.~-G.~Huang and Y.~Wang,
  ``Large local non-Gaussianity from general ultra slow-roll inflation,''
  JCAP {\bf 1306}, 035 (2013).

\bibitem{Chen:2013kta}
  X.~Chen, H.~Firouzjahi, M.~H.~Namjoo and M.~Sasaki,
  ``Fluid Inflation,''
  arXiv:1306.2901 [hep-th].


\bibitem{Ade:2013ydc} 
  P.~A.~R.~Ade {\it et al.}  [Planck Collaboration],
  ``Planck 2013 Results. XXIV. Constraints on primordial non-Gaussianity,''
  arXiv:1303.5084 [astro-ph.CO].





\bibitem{Weinberg:2005vy}
  S.~Weinberg,
  ``Quantum contributions to cosmological correlations,''
  Phys.\ Rev.\ D {\bf 72}, 043514 (2005)
  [hep-th/0506236].

\bibitem{Chen:2010xka}
  X.~Chen,
  ``Primordial Non-Gaussianities from Inflation Models,''  Adv.\ Astron.\  {\bf 2010}, 638979 (2010)  [arXiv:1002.1416 [astro-ph.CO]].  

\bibitem{Chen:2009zp}
  X.~Chen and Y.~Wang,
  ``Quasi-Single Field Inflation and Non-Gaussianities,''
  JCAP {\bf 1004}, 027 (2010)
  [arXiv:0911.3380 [hep-th]].
  
\bibitem{Lyth:2013sha} 
  D.~H.~Lyth and M.~Karciauskas,
  ``The statistically anisotropic curvature perturbation generated by $f(\phi)^2 F^2$,''
  JCAP {\bf 1305}, 011 (2013)
  [arXiv:1302.7304 [astro-ph.CO]].
  
\bibitem{Jain:2012vm} 
  R.~K.~Jain and M.~S.~Sloth,
  ``On the non-Gaussian correlation of the primordial curvature perturbation with vector fields,''
  JCAP {\bf 1302}, 003 (2013)
  [arXiv:1210.3461 [astro-ph.CO]].

  
\bibitem{Nurmi:2013gpa} 
  S.~Nurmi and M.~S.~Sloth,
  ``Constraints on Gauge Field Production during Inflation,''
  arXiv:1312.4946 [astro-ph.CO].


\bibitem{Ohashi:2013qba} 
  J.~Ohashi, J.~Soda and S.~Tsujikawa,
  ``Observational signatures of anisotropic inflationary models,''
  JCAP {\bf 1312}, 009 (2013)
  [arXiv:1308.4488 [astro-ph.CO], arXiv:1308.4488].

\bibitem{Fujita:2013qxa} 
  T.~Fujita and S.~Yokoyama,
  ``Higher order statistics of curvature perturbations in IFF model and its Planck constraints,''
  JCAP {\bf 1309}, 009 (2013)
  [arXiv:1306.2992 [astro-ph.CO]].

\bibitem{Shiraishi:2013sv} 
  M.~Shiraishi, S.~Yokoyama, K.~Ichiki and T.~Matsubara,
  ``Scale-dependent bias due to primordial vector field,''
  arXiv:1301.2778 [astro-ph.CO].

\bibitem{Baghram:2013lxa} 
  S.~Baghram, M.~H.~Namjoo and H.~Firouzjahi,
  ``Large Scale Anisotropic Bias from Primordial non-Gaussianity,''
  JCAP {\bf 1308}, 048 (2013)
  [arXiv:1303.4368 [astro-ph.CO]].

\bibitem{Thorsrud:2013kya} 
  M.~Thorsrud, F.~R.~Urban and D.~F.~Mota,
  ``Statistics of Anisotropies in Inflation with Spectator Vector Fields,''
  arXiv:1312.7491 [astro-ph.CO].

\bibitem{Rodriguez:2013cj} 
  Y.~Rodriguez, J.~P.~Beltran Almeida and C.~A.~Valenzuela-Toledo,
  ``The different varieties of the Suyama-Yamaguchi consistency relation and its violation as a signal of statistical inhomogeneity,''
  JCAP {\bf 1304}, 039 (2013)
  [arXiv:1301.5843 [astro-ph.CO]].

\bibitem{BeltranAlmeida:2011db} 
  J.~P.~Beltran Almeida, Y.~Rodriguez and C.~A.~Valenzuela-Toledo,
  ``The Suyama-Yamaguchi consistency relation in the presence of vector fields,''
  Mod.\ Phys.\ Lett.\ A {\bf 28}, 1350012 (2013)
  [arXiv:1112.6149 [astro-ph.CO]].

\bibitem{Shiraishi:2013oqa} 
  M.~Shiraishi, E.~Komatsu and M.~Peloso,
  ``Signatures of anisotropic sources in the trispectrum of the cosmic microwave background,''
  arXiv:1312.5221 [astro-ph.CO].






\end{thebibliography}
\end{document}